\documentclass[preprint,12pt,amsmath,amssymb,showkeys]{revtex4} 
\usepackage{amsmath}
\usepackage{latexsym}
\usepackage{amssymb}
\usepackage{graphics,epstopdf}
\usepackage{amsmath,amssymb,amsthm}
\theoremstyle{plain}
\newtheorem{thm}{Theorem}
\newtheorem{proposition}{Proposition}
\usepackage{graphicx}
\usepackage{enumerate}
\usepackage[colorlinks=true, citecolor=blue, urlcolor=blue ]{hyperref}

\begin{document}
\title{ Factorization, partner Hamiltonians and intelligent states in $\mathcal{P}\mathcal{T}$-symmetric position dependent effective mass systems }

\author{Pinaki Patra}
\thanks{Corresponding author}
\email{monk.ju@gmail.com}
\affiliation{Department of Physics, Brahmananda Keshab Chandra College, Kolkata, India-700108}

\date{\today}

\begin{abstract}
We study a factorized $\mathcal{P}\mathcal{T}$-symmetric position-dependent mass Hamiltonian on a finite interval with quasiperiodic boundary conditions. A positive, bounded, and boundedly invertible metric is constructed explicitly, and the Hamiltonian is shown to be self-adjoint in the associated metric Hilbert space and similar to a Hermitian position-dependent-mass operator. Consequently, its spectrum is real within the stated parameter domain. The reverse factorization defines a $\mathcal{P}\mathcal{T}$-symmetric partner Hamiltonian, and the first-order factors act as intertwining operators.  Eigenstates of the lowering-type factorization operator are consequently interpreted as Robertson intelligent states, not as states generated by an energy ladder. Their admissible eigenvalues are quantized by the quasiperiodic boundary condition, and they saturate the corresponding Robertson-Schr\"{o}dinger uncertainty relation. A weakly position-dependent mass profile illustrates the construction within the explicitly quantified regime.
\end{abstract}
 
 \maketitle
\section{Introduction}
 Unitarity of the time-evolution operator and real eigenvalues of quantum observables are two fundamental cornerstones of quantum mechanics.  For a self-adjoint Hamiltonian ($\hat{H}$), the condition of the real spectrum is trivially followed, whereas the evolution operator is given by $\hat{U}(t)=e^{-i\hat{H}t/\hbar}$ \cite{QM1}. Nonetheless, non-Hermitian Hamiltonians are still useful  for
studying open quantum systems in nuclear physics or
quantum optics, among others \cite{pt1,pt2}. These non-Hermitian Hamiltonians are considered an effective subsystem within a projective subspace of the total system, which obeys conventional quantum mechanics with a Hermitian Hamiltonian \cite{pt1,pt2,Lee1}.  However, in 1998, Bender et al. had shown that the issues of the unitarity and real energy eigenvalues can be taken into account with the help of a weaker condition on the Hamiltonian, namely parity-time ($\mathcal{P}\mathcal{T}$)- symmetric Hamiltonians, along with a deformed inner product \cite{bender1,bender2,CPTconjugation}.  The space-reflection linear operator, namely the parity operator ($\hat{\mathcal{P}}$), changes the sign of the position ($\hat{x}$) and momentum ($\hat{p}$) operators ($\hat{\mathcal{P}}\hat{x}\hat{\mathcal{P}}^{-1}=-\hat{x}$, $\hat{\mathcal{P}}\hat{p}\hat{\mathcal{P}}^{-1}=-\hat{p}$), keeping the fundamental commutation relation (the Heisenberg algebra $[\hat{x},\hat{p}]=i\hbar$) invariant. From now on we shall consider the Planck's constsnt $\hbar=1$, unless otherwise specified. The anti-unitary ($\hat{\mathcal{T}}i \hat{\mathcal{T}}^{-1}=-i$) time-reversal operator ($\hat{\mathcal{T}}$) changes the sign of momentum  ($\hat{\mathcal{T}}\hat{p}\hat{\mathcal{T}}^{-1}=-\hat{p}$), but leaves $\hat{x}$ invariant ($\hat{\mathcal{T}}\hat{x}\hat{\mathcal{T}}^{-1}=\hat{x}$), keeping the Heisenberg-algebra intact. Since, $\hat{\mathcal{P}}$ and $\hat{\mathcal{T}}$ are similar to the reﬂection operators, they are involutory operators ($\hat{\mathcal{P}}^2=\hat{\mathcal{T}}^2 =\hat{\mathbb{I}}$, where $\hat{\mathbb{I}}$ is the identity operator). Moreover, $\hat{\mathcal{P}}$ and $\hat{\mathcal{T}}$ commutes with each other. We say that  $\hat{H}$ is $\mathcal{P}\mathcal{T} $-symmetric if $[\hat{H},\hat{\mathcal{P}}\hat{\mathcal{T}}]=0 $. \\
If the real spectrum and the unitarity are the concerns, then $\mathcal{P}\mathcal{T} $-symmetric Hamiltonians had opened up the possibilities of incorporating a larger class of operators to be considered as a viable Hamiltonian for a quantum mechanical system. A wide class of systems had been studied for the last two decades \cite{pt3,pt4,pt5,pt6,pt7,pt8,pt80}. For an operational foundations of $\mathcal{P}\mathcal{T} $-symmetric and quasi-Hermitian quantum theory, one can see \cite{pt7,pt8,pt9,pt10}. \\
 Complementary, another important issue in quantum mechanics is the concept of the position-dependent effective mass (PDEM), which was incepted to describe the electronic properties and band structure of semiconductor Physics \cite{heterostructure1,heterostructure2}. Later, it was proved to be exist for various domain of Physics. Notably, frequent occurrence of effective mass in the study of nonlinear optical properties in quantum well, the asymmetric shape of crackling noise pulses emitted by a diverse range of noisy systems, in cosmological models, and even in quantum information theory, had made PDEM a topical issue \cite{pdemapplication1,pdemapplication2,pdemapplication3,pdemapplication4,pdemapplication5,pdemapplication6,pdemapplication7,pdemapplication8,pdemapplication9}. Since PDEM and momentum do not commute with each other, there is ambiguity in the form of a viable kinetic part of PDEM Hamiltonian \cite{pdemambiguity1,pdemambiguity2,pdemambiguity3,pdemambiguity4}. We shall consider a fairly accepted form of PDEM Hamiltonian in the present paper \cite{pdemkineticpart1,pdemkineticpart2,pdemkineticpart3,pdemkineticpart4}. \\
 Intelligent states (IS) in quantum mechanics is a specific type of quantum state that minimizes the Robertson-Scr\"{o}dinger uncertainty relation to an equality (achieving the theoretical minimum possible uncertainty for two given observables) \cite{is1,is2,is3,is4,is5,is6}. Since IS can be termed as a generalized coherent state (CS), we shall often interchange the terminology CS and IS in the present work \cite{is1,is2,is3,is4,is5,is6,cs1,cs2,cs3,cs4}. In the present paper, we factorize the Hamiltonians in terms of first order differential operators $\hat{A}_{\pm}$ \cite{factorization1,factorization2,factorization3}. However, in the present context, since $[\hat{A}_{-},\hat{A}_{+}]=\theta(x)$ is not constant, we can not interpret them as pair of annihilation and creation operators \cite{bogdan}. We  call $\hat{A}_{\pm}$ as the $\eta$-adjoint factorization operators.  Nonetheless, $\theta(x)$ becomes contant for Hermitian limit. This suggest us to interpret the right-eigenstates   of $\hat{A}_{-}$ as the generalized intelligent states (IS). Then, we define generalized phase-space operators ($\hat{X}$ and $\hat{\Pi}$), and construct the IS $\psi_{\nu}$ through the eigenvalue equation $(\hat{X}+i\hat{\Pi})\psi_{\nu}=\sqrt{2}\nu \psi_{\nu}$ for a Robertson IS of the pair  ($\hat{X},\hat{\Pi}$).    It is worth noting that, the factorization method we utilize here is a generalization of factorization method 
 for a Hermitian Hamiltonian, in which the  factorization of the Hamiltonian can be done with the help of intertwining operator which connects the Hamiltonian with its supersymmetric (SUSY) partner \cite{bogdan,susy1,susy2,susy3,susy4}.  IS, and CS structures for PDEM systems are well studied in the literature \cite{is1,is2,is3,is4,is5,is6,susypdem1,susypdem2,susypdem3}. The $\mathcal{P}\mathcal{T}$ symmetric quantum system for PDEM is studied in the literature in \cite{ptpdem1,ptpdem2,ptpdem3,ptpdem4,ptpdem5,ptpdem6,ptpdem7,ptpdem8}. In the present paper, we have shown that for $\mathcal{P}\mathcal{T}$-symmetric case, the intertwining operators are not sufficient to factorize the Hamiltonian. We have factorized our Hamiltonian under consideration with the help of an ansatz, which is a deformed version of the intertwining operator connecting the SUSY partner of the original Hamiltonian. It turns out that the factorizing operators satisfy the algebra similar to  the annihilation and creation operators in a deformed space (deformed position and momentum).  We explore the general scope of uncertainty principle for the concerned $\mathcal{P}\mathcal{T}$-symmetric scenario. The explicit computation for a toy model is outlined to support our claims.\\
The organization of the paper is  followings. At first we have described the system under consideration and the choice of a viable inner product for our system. Then  SUSY formalism for  our system is outlined, along with the factorization of the Hamiltonian, we explicitly discuss the domains of the differential operators under consideration.  After that, we construct the generalized intelligent state  structure and discuss the uncertainty relation of the generalized quadrature in the deformed space.    At last, we provide a quantitative analysis for a toy model upto first order in the $\mathcal{P}\mathcal{T}$ symmetric parameter. 
 \section{System under consideration}
Let us consider a $\mathcal{P}\mathcal{T}$-symmetric generalized parametric amplifier, with amplification parameters $\beta_1$ and $\beta_2$ and with position dependent effective mass (PDEM) $m(x)$ is placed in an one dimension box ($x\in (-L,L)$) of length $2L$. Hamiltonian ($\hat{H}$) for this generalized Swanson-system is given by
\begin{equation}\label{hamiltonian}
 \hat{H}= \hat{p}\frac{1}{m}\hat{p} -\frac{1}{2m}\hat{p}m\hat{p}\frac{1}{m} + V(x) + i (\beta_1 \hat{p}\hat{x} + \beta_2 \hat{x}\hat{p}).
\end{equation}
For constant $m$ and the absence of the $\mathcal{P}\mathcal{T}$-symmetric term (dependent on the parameters $\beta_1$ and $\beta_2$) will lead to the usual Hermitian Hamiltonian. For convenience, we shall consider the following sufficient conditions to ~\eqref{hamiltonian} be  $\mathcal{P}\mathcal{T}$-symmetric.
\begin{enumerate}[i.]
	\item $\beta_1, \beta_2 \in \mathbb{R}$.
	\item If $V(x): \mathbb{R}\to \mathbb{R}$, then $V(x)$ is an even function ($V(-x)=V(x)$). \\
	Otherwise, $V(x)=iV_0(x)$, with $V_0(x): \mathbb{R}\to \mathbb{R}$, and $V_0(x)$ is an odd function ($V_0(-x)=-V_0(x)$).
	\item If $m(x): \mathbb{R}\to \mathbb{R}$, then $m(-x)=m(x)$. \\
	Otherwise, $m(x)=im_0(x)$, with $m_0(x): \mathbb{R}\to \mathbb{R}$, and $m_0(x)$ is an odd function.
\end{enumerate}
For our present study, we consider the external potential as
\begin{equation}\label{Vx}
	V(x)=V_0+kx^2,\; V_0\in \mathbb{R},\; k\ge 0.
\end{equation}
Note that, in a practical experiment,  experimenters can tune the potential well ($V_0<0$) or barrier ($V_0>0$), the oscillator parameter $k$, amplification parameters $\beta_1$, $\beta_2$, and the box length $2L$.
In the position representation ($\{\vert x\rangle\}$), the Hamiltonian  ~\eqref{hamiltonian} reads
\begin{equation}\label{Hposition}
	\hat{H}=-\frac{1}{2m}\frac{\partial^2}{\partial x^2} + u(x)\frac{\partial}{\partial x} + V_{e}(x),
\end{equation}
with
\begin{eqnarray}
	u(x)=(\beta_1+\beta_2)x -\frac{1}{2}\left(\frac{1}{m}\right)',\\
	V_{e}(x)=\beta_1 -\frac{1}{2m}\left(\frac{m'}{m}\right)' + V(x).
\end{eqnarray}
Here prime $(')$ represents derivative with respect to $x$.
 The adjoint ($\hat{H}^\dagger$) of $\hat{H}$ is given by
\begin{equation}\label{hdagger}
 \hat{H}^\dagger= \hat{p}\frac{1}{m}\hat{p} -\frac{1}{2m}\hat{p}m\hat{p}\frac{1}{m} + V(x) - i (\beta_1 \hat{x}\hat{p} + \beta_2 \hat{p}\hat{x}),
\end{equation}
which is not equal to $\hat{H}$ for $\beta_1\neq -\beta_2$. In other words, $\hat{H}$ is Hermitian iff $\beta_1=-\beta_2$. \\
 If an invertible Hermitian operator $ \hat{\eta}: \mathcal{H}\to \mathcal{H}$ exists, such that 
\begin{equation}\label{etahhdaggerconnection}
\hat{H}^\dagger = \hat{\eta}\hat{H}\hat{\eta}^{-1},
\end{equation}
then the linear operator $ \hat{H}: \mathcal{H}\to \mathcal{H}$ is said to be pseudo-Hermitian. 
To see $\hat{H}$ in ~\eqref{hamiltonian} is a pseudo-Hermitian operator, let us show the existence of a metric operator $\hat{\eta}$ such that ~\eqref{etahhdaggerconnection} holds.
Let us use the ansatz 
\begin{equation}
 \hat{\eta}= e^{\hat{\Lambda}(x)}
\end{equation}
in ~\eqref{etahhdaggerconnection} and use
 the Baker-Campbell-Hausdorff formula. We get the following set of equations.
\begin{eqnarray}
e^{\hat{\Lambda}}\hat{p}\frac{1}{m}\hat{p} e^{-\hat{\Lambda}} &=& \hat{p}\frac{1}{m}\hat{p} + i \{\hat{p}, \frac{1}{m}\Lambda' \} -\frac{1}{m}(\Lambda')^2.\\
e^{\hat{\Lambda}} \frac{1}{m} \hat{p} m \hat{p} \frac{1}{m} e^{-\hat{\Lambda}} &=& \frac{1}{m} \hat{p} m \hat{p} \frac{1}{m} + i (\Lambda' \hat{p}\frac{1}{m} + \frac{1}{m} \hat{p} \Lambda') +\frac{i}{m}(\Lambda')^2 .\\
e^{\hat{\Lambda}} V(x) e^{-\hat{\Lambda}} &=& V(x).\\
e^{\hat{\Lambda}} i(\beta_1 \hat{p}\hat{x} + \beta_2 \hat{x}\hat{p}) e^{-\hat{\Lambda}} &=& i (\beta_1 \hat{p}\hat{x} + \beta_2 \hat{x}\hat{p}) - \beta_{12} x \Lambda' , \;\mbox{with}\; \beta_{12}=\beta_1+\beta_2.
\end{eqnarray}
Here prime denotes the derivative with respect to $x$, and $\{u,v\}:= uv+vu$ denotes the anti-commutator of $u$ and $v$.
Now we can write the following transformation law of the Hamiltonian.
\begin{equation}\label{etahetainverse}
 e^{\hat{\Lambda}} \hat{H} e^{-\hat{\Lambda}} = \hat{H} + \frac{i}{2} \{\hat{p}, \frac{1}{m}\Lambda' \} -\frac{1}{2m}(\Lambda')^2 -  \beta_{12} x \Lambda' .
\end{equation}
Comparing the equations ~\eqref{hdagger}, ~\eqref{etahhdaggerconnection} and ~\eqref{etahetainverse}, we arrive at the following sufficient condition for $\Lambda(x)$.
\begin{equation}\label{lambdaprime}
 \Lambda'(x) = -2\beta_{12}x m(x).
\end{equation}
The first order equation ~\eqref{lambdaprime} always admits a solution for all locally integrable $m(x)$. For $\beta_1=-\beta_2$, $\hat{H}$ is identical with its adjoint for the choice of integration constant of the solution of ~\eqref{lambdaprime} to be zero, which will be assumed throughout this paper.
In particular, the Hamiltonian ~\eqref{hamiltonian} is pseudo-Hermitian, and $\hat{H}$  is connected with its adjoint by the similarity transformation
\begin{equation}\label{eta}
\hat{\eta} = \exp \left[ -2\beta_{12}\int^x ym(y)dy \right].
\end{equation}
For a continuous $m(x)>0$ and $m(-x)=m(x)$, we have
\begin{equation}
	0\le F(x)\le F_L,\; \mbox{where}\; F_L=\int_{0}^{L}ym(y)dy<\infty,
\end{equation}
which implies
\begin{equation}
	e^{-2\beta_{12}F_L} \le \eta(x)\le e^{2\beta_{12}F_L},\; \mbox{and}\; 
	e^{-2\beta_{12}F_L} \le \eta^{-1}(x)\le e^{2\beta_{12}F_L}.
\end{equation}
Therefore for $\eta(x)>0$, we can state the following proposition.
\begin{proposition}
Both	$\eta$ and $\eta^{-1}$ are bounded.
\end{proposition}
Existence of $\hat{\eta}$ suggests the following inner  product
\begin{equation}\label{innerproductmod}
 \langle \zeta_1 \vert \zeta_2 \rangle = (\zeta_1,\hat{\eta}\zeta_2),
\end{equation}
where $(u,v)$ is the usual inner-product in $\mathbb{L}^2 $ space. Moreover, the metric norm and $\mathbb{L}^2$ norm are equivalent:
\begin{equation}
e^{-\beta_{12}F_L}\vert\vert \psi\vert\vert_{\mathbb{L}^2} \le \vert\vert\psi\vert\vert_{\eta} \le e^{\beta_{12}F_L} \vert\vert\psi\vert\vert_{\mathbb{L}^2}.
\end{equation}
 We shall use the inner product ~\eqref{innerproductmod} in the subsequent sections. In particular, we shall work in the metric Hilbert space $\mathcal{H}_{\eta}=\mathbb{L}^2((-L,L),\eta(x)dx)$.
\section{Factorization of Hamiltonian}
Since we are dealing with a non-hermitian Hamiltonian ($\hat{H}$), the intertwining operator $\hat{A}$ will not be sufficient to factorize $\hat{H}$. Let us consider the following factorization of $H$ 
\begin{equation}\label{Hfact}
	\hat{H} = \hat{A}_{+}\hat{A}_{-}+\lambda.
\end{equation}
with ansatz
\begin{equation}\label{ansatzAplummiinuus}
 \hat{A}_{\pm} = \mp \alpha(x)\frac{d}{dx}\alpha(x) +\phi_{\pm}(x).
\end{equation}
Comparing the coefficients of $\psi$ from $\hat{H}\psi$ for  ~\eqref{Hfact} and ~\eqref{Hposition}, we get following consistency conditions. The coefficients of $\psi''$ gives the following form of $\alpha(x)$.
\begin{equation}\label{alphax}
	\alpha(x)=(2m)^{-1/4}.
\end{equation}
Since we consider $m(-x)=m(x)$, we also have $\alpha(-x)=\alpha(x)$.
Coefficients of $\psi'$ provides the relation between $\phi_{\pm}(x)$ as
\begin{equation}\label{phipmdifference}
	\phi_{+}-\phi_{-}=4\alpha\alpha'+\frac{u}{\alpha^2} = \beta_{12}x\sqrt{2m}:= u_0(x).
\end{equation}
Note that,  $4\alpha\alpha'+u/\alpha^2=0$ implies $\beta_1=-\beta_2$, which corresponds to a Hermitian counterpart of our concerned system. In particular, for hermitian system $\phi_+(x)=\phi_{-}(x)$. Since we emphasize in non-hermitian $\mathcal{P}\mathcal{T}$ symmetric system in the present paper, we shall consider  
\begin{equation}
4\alpha\alpha'+u/\alpha^2\neq 0 \implies \beta_{12}\neq 0,\; \phi_+\neq \phi_{-}.
\end{equation}
The coefficients of $\psi$ provides the following Riccati equation for $\phi_{-}(x)$.
\begin{eqnarray}\label{ricatti}
	\phi_{-}' -\sqrt{2m}\phi_{-}^2 - 2m\beta_{12}x\phi_{-} = \sqrt{2m}(\lambda-V_{e}) - \frac{\beta_{12}x}{2\sqrt{2}}\left(\frac{m'}{m}\right) \nonumber\\ + \frac{1}{4\sqrt{2m}}\left(\frac{m''}{m}-\frac{7}{4}\left(\frac{m'}{m}\right)^2\right),
\end{eqnarray}
which means, given the PDEM $m(x)$ one can determine the superpotential $\phi_(x)$ and construct required $\hat{A}_\pm$. The commutation relation between $\hat{A}_{-}$ and $\hat{A}_{+}$ reads
\begin{equation}\label{Apmcommut}
	[\hat{A}_{-},\hat{A}_+]=\theta(x), \;\mbox{with}\; \theta= \frac{1}{\sqrt{2m}}(\phi_{-}+\phi_{+})'.
\end{equation}
The intertwining partner Hamiltonian $\hat{H}_1$ of $\hat{H}$ may be defined as
\begin{equation}
	\hat{A}_{-}\hat{H}=\hat{H}_1\hat{A}_{-},\; \mbox{and}\; \hat{A}_{+}\hat{H}_1=\hat{H}\hat{A}_{+}.
\end{equation} 
Using ~\eqref{Apmcommut}, and the explicit form ~\eqref{Hposition}, one can write
\begin{equation}
	\hat{H}_1=\hat{H}+\theta(x).
\end{equation}
In the Hilbert space $\mathcal{H}_{\eta}$, we consider the domain as a common dense core $\mathcal{C}_{\mathcal{H}_\eta}=C^\infty(-L,L)$ of the unbounded differential operators $\hat{A}_{\mp}, \hat{H},\hat{H}_1$. Under the metric inner product $\langle f,g\rangle_{\eta}=\int_{-L}^{L}f^*(x)g(x)\eta(x)dx$, using integration by parts, we see that
\begin{equation}\label{adjointdefn}
	\langle f,\hat{A}_{-}g\rangle_{\eta}-\langle \hat{A}_{+}f,g\rangle_{\eta}=\Lambda_{L}= \Lambda_{f,g}(L)-\Lambda_{f,g}(-L),\;\mbox{with}\; \Lambda_{f,g}(x)= f^*g\alpha^2\eta.
\end{equation}
Here $*$ denotes complex conjugation.
If the boundary term $\Lambda_L$ vanishes, then the $\eta$-adjoint ($(A^\dagger)^\dagger=\eta^{-1}A^\dagger  \eta$) of $\hat{A}_{-}$ is given by
\begin{eqnarray}\label{Apmconnection}
	(A_{-}^\dagger)^\dagger= \eta^{-1}A_{-}^\dagger  \eta = -\alpha \frac{d}{dx}\alpha +\phi_{-}-\alpha^2 \Lambda' =A_{+},
\end{eqnarray}
where we have used ~\eqref{lambdaprime} and ~\eqref{phipmdifference}. Let us choose a quasiperiodic domain for $\hat{A}_{-}$ by 
\begin{equation}
	\mathcal{D}_{\chi}(\hat{A}_{-})= \{ \psi\in \mathcal{H}_{\eta}: \alpha\psi \in AC[-L,L], \hat{A}_{-}\psi \in \mathcal{H}_{\eta}, \psi(L)=e^{i\chi}\psi(-L)\},\; \mbox{with}\; \chi\in[0,2\pi].
\end{equation}
Note that $\chi=0$ corresponds to periodic boundary condition (BC). Similarly for $\hat{A}_{+}$ we write
\begin{equation}
	\mathcal{D}_{\chi}(\hat{A}_{+})= \{ \psi\in \mathcal{H}_{\eta}: \alpha\psi \in AC[-L,L], \hat{A}_{+}\psi \in \mathcal{H}_{\eta}, \psi(L)=e^{i\chi}\psi(-L)\},\; \mbox{with}\; \chi\in[0,2\pi].
\end{equation}
The notation $AC[-L,L]$ stands for absolute continuous  functions on $[-L,L]$. Since $\eta$ and $\alpha$ are even functions of $x$, under the quasiperiodic BC ($f(L)=e^{i\chi}f(-L)$ , $g(L)=e^{i\chi}g(-L)$), we note that the boundary term of ~\eqref{adjointdefn} vanishes as follows.
\begin{equation}
	\Lambda_{L}=f^*(L)g(L)\alpha^2(L)\eta(L)-f^*(-L)g(-L)\alpha^2(-L)\eta(-L) =0
\end{equation}
What follows is that $\hat{A}_{\pm}$ form a genuine $\eta$-adjoint pair on $\mathcal{D}_{\chi}(\hat{A}_{\pm})$. Moreover, we have the natural domain of the factorized Hamiltonian and its partner Hamiltonian as
\begin{equation}\label{domainHH1}
	\mathcal{D}(\hat{H})=\{\psi\in \mathcal{D}_{\chi}(\hat{A}_{-}): \hat{A}_{-}\psi \in \mathcal{D}_{\chi}(\hat{A}_{+})\},
	\mathcal{D}(\hat{H}_1)=\{\psi\in \mathcal{D}_{\chi}(\hat{A}_{+}): \hat{A}_{+}\psi \in \mathcal{D}_{\chi}(\hat{A}_{-})\}.
	\end{equation}
Thus a function in $\mathcal{D}(\hat{H})$ must satisfy the quasiperiodic condition 
\begin{equation}\label{bc}
\psi(L)=e^{i\chi}\psi(-L).
\end{equation}
Using ~\eqref{Apmconnection}, we can say that $\hat{N}=\hat{A}_{+}\hat{A}_{-}$ is a nonnegative self-adjoint operator in $\mathcal{H}_{\eta}$. Accordingly, $\hat{H}=\hat{N}+\lambda$ is also self-adjoint in $\mathcal{H}_{\eta}$ for real $\lambda$. Now, for every $\psi\in \mathcal{D}(\hat{H})$, 
\begin{equation}
	(\psi,\hat{H}\psi)_{\eta}=(\psi, (\hat{A}_{-}^\dagger)^\dagger \hat{A}_{-}\psi)_\eta +\lambda \vert \vert\psi\vert\vert_{\eta}^2 = \vert \vert \hat{A}_{-}\psi\vert\vert^2_{\eta} +\lambda \vert\vert\psi\vert\vert^2_{\eta}.
\end{equation}
In other words, the spectrun of $\hat{H}$, $\sigma(\hat{H})\subseteq [\lambda,\infty) \subset \mathbb{R}$ enables us to write the following proposition.
\begin{proposition}
	Specrum of $\hat{H}$ is real in $\mathcal{H}_{\eta}$.
\end{proposition}
Using the explicit form ~\eqref{eta} and ~\eqref{Hposition}, we can write an equivalent self-adjoint Hamiltonian
\begin{equation}
	\hat{h}=\hat{\rho}\hat{H}\hat{\rho}^{-1}=-\frac{d}{dx}\left[\frac{1}{2m(x)}\frac{d}{dx}\right]+ V_{equiv}(x),\;\; \hat{\rho}=\hat{\eta}^{1/2},
\end{equation}
where the equivalent potential reads
\begin{equation}
	V_{equiv}= V_{e}(x) +\frac{1}{2}\beta_{12}^2 m(x)x^2 -\frac{1}{2}\beta_{12}.
\end{equation}
\section{Intelligent state and uncertainty relation}
The algebra ~\eqref{Apmcommut} of $\hat{A}_{\mp}$ indicates, one can define a deformed phase-space with co-ordinate $\hat{X}$ and momentum $\hat{\Pi}$ as 
\begin{equation}\label{XPidefn}
	\hat{X}=\frac{1}{\sqrt{2}}(\hat{A}_{+}+\hat{A}_{-}),\; 
	\hat{\Pi}=\frac{i}{\sqrt{2}}(\hat{A}_{+}-\hat{A}_{-}),
\end{equation}
which admits the commutation relation
\begin{equation}\label{XPicommutator}
	[\hat{X},\hat{\Pi}]=i \theta.
\end{equation}
From the definition ~\eqref{ansatzAplummiinuus}, we see that $X=(\phi_{+}(x)+\phi_{-}(x))/\sqrt{2}$ is an ordinary multiplication operator. Moreover, from ~\eqref{Apmconnection}, we have $(\hat{X}^\dagger)^\dagger=\hat{X}$, which implies $X$ is a bounded self-adjoint operator on the entire metric Hilbert space. To see the self-adjointness of the momentum like quadrature $\Pi$, wee define the unitary map
\begin{equation}
	(U\psi)(q)=\alpha(x(q))\sqrt{\eta(x(q))}\psi(x(q)),\;\mbox{where}\; q(x)=\int^x \sqrt{2m(s)}ds.
\end{equation}
Now we can write
\begin{equation}
	\int \vert U\psi(q)\vert^2 dq =\int \eta(x)\vert \psi(x)\vert^2 dx\;\; (\because \; dq=\alpha^{-2}(x)dx).
\end{equation}
Using ~\eqref{phipmdifference} and ~\eqref{lambdaprime}, we have
\begin{equation}
	U\Pi U^{-1}=-i\sqrt{2}\frac{d}{dq},
\end{equation}
which is self-adjoint on the quasiperiodic domain. Hence, we have the consistent $\eta$-adjoint situation $(\hat{\Pi}^\dagger)^\dagger=\hat{\Pi}$. Hence, $X$ and $\Pi$ are valid observable under consideration.
On the other hand, we see that the operator $\hat{N}=\hat{A}_{+}\hat{A}_{-}$, satisfy the algebra
\begin{equation}
	[\hat{N},\hat{A}_{+}]= \hat{A}_{+}\theta(x),\; [\hat{N},\hat{A}_{-}]= -\theta(x)\hat{A}_{-}.
\end{equation}
We see that even if for some eigenstate state $\vert n\rangle$ with eigenvalue $n$ of $\hat{N}$, we can not interpret $\hat{A}_{\pm}\vert n\rangle$  as an eigenvector of $\hat{N}$, with eigenvalues $n\pm \theta$ (since $\theta(x)$ is an operator here). In other words, for general $\theta(x)$, we can only interpret $\hat{A}_{\pm}$ as factorization operators or metric-adjoint pair of factorization and intertwining operators.
Note that for constant $\theta$, $\hat{A}_{\pm}$ satisfy the algebra for annihilation and creation operators. 
The Riccati equation ~\eqref{ricatti}, contains an undetermined parameter $\lambda$. To fix its value, let us recall that, in constant mass oscillator, it is proportional to the commutator of the annihilation and creation operators.  
Keeping in mind the terminology of Hermitian quantum system, we define the generalized intelligent state ($\vert\nu\rangle_R$), as the right eigenstate of  $\hat{A}_{-}$. In particular
 \begin{equation}
	\hat{A}_{-}\vert \nu\rangle_r =\nu \vert \nu\rangle_r ,\; \nu=\nu_r+i\nu_c, \mbox{with}\; \nu_r,\nu_c\in \mathbb{R}.
\end{equation}
Note that left eigenstate $_l\langle \nu \vert$ of $\hat{A}_{-}$ can be obtained from the right eigenstate of $\hat{A}_{-}^\dagger$, where $\hat{A}_{-}^\dagger$ is the adjoint of $\hat{A}_{-}$. In particular,
\begin{equation}
\hat{A}_{-}^\dagger \vert \nu\rangle_l =\nu\vert \nu\rangle_l.
\end{equation}
Let $\psi_{r\nu}(x)$ is a right eigenstate of $\hat{A}_{-}$ in position representation and $\psi_{l\nu}(x)$ is a  left eigenstate of $\hat{A}_{+}$. We note that $\psi_{l\nu}(x)$ is a right eigenstate of $\hat{A}_{+}^\dagger$. In other words
\begin{eqnarray}
	\alpha\frac{d}{dx}(\alpha\psi_{r\nu})+\phi_{-}\psi_{r\nu}=\nu \psi_{r\nu}, \label{Aminuseigen}\\
		\alpha\frac{d}{dx}(\alpha\psi_{l\nu})+\phi_{+}\psi_{l\nu}=\nu \psi_{l\nu}. \label{Apluseigen}
\end{eqnarray}
Using explicit form of $\alpha(x)$, relation between $\phi_{\pm}(x)$ and explicit form of $\eta(x)$ from equations \eqref{alphax}, \eqref{phipmdifference} and \eqref{eta}, respectively, one can identify that
\begin{equation}
	\psi_{l\nu}(x)= \eta(x)\psi_{r\nu}(x),
\end{equation}
which enables us to write down the following theorem.
\begin{thm}\label{innerproducttheorem}
For a $\mathcal{P}\mathcal{T}$ symmetric position-dependent effective mass system defined by \eqref{hamiltonian}, which admits factorization of the form \eqref{Hfact} through \eqref{ansatzAplummiinuus}, the $\mathbb{L}^2$ inner product $( \psi_{l}, \psi_{r})$ is equivalent to $( \psi_{r}, \eta \psi_{r})$, where $\psi_r$ is a right eigenstate and $\psi_l$ is a left eigenstate of $\hat{A}_{-}$ and $\hat{A}_{+}$, respectively.
\end{thm}
Note that the Theorem-~\ref{innerproducttheorem} is independent to the form of  external potential $V(x)$. Now solving ~\eqref{Aminuseigen}, we write the explicit form of the generalized IS (GIS)
\begin{equation}
	\psi_{r\nu}(x)=\mathcal{N}_{\nu}\sqrt[4]{2m(x)}e^{\int^x \sqrt{2m(s)}(\nu- \phi_{-}(s))ds},
\end{equation}
where the normalization constant $\mathcal{N}_{\nu}$ is given by the condition
\begin{equation}
	\vert \mathcal{N}_{\nu}\vert^2 \int_{-L}^{+L} \sqrt{2m(x)} e^{2\Re[\int^x \sqrt{2m(s)}(\nu- \phi_{-}(s))ds]}\eta(x)dx =1.
\end{equation}
The notations $\Re[z]$ and $\Im[z]$ stand for real and imaginary part of $z$. Note that the state $\psi_{r0}$ corresponding to $\nu=0$, may be interpreted as the vacuum state. One can write a GIS in terms of vacuum as 
\begin{equation}\label{psirnupsir0relation}
	\psi_{r\nu}(x)=(\mathcal{N}_{\nu}/\mathcal{N}_{0})e^{\nu q(x)}\psi_{r0}(x),\; \mbox{with}\; q(x)=\int^x \sqrt{2m(s)}ds.
\end{equation}
Using the BC ~\eqref{bc}, we see that the allowed eigen-values $\nu$ of the factorization operator $\hat{A}_{-}$ is given by
\begin{equation}\label{nunallowed}
	\nu_n= i(n\pi +\chi/2)/q(L),\; \mbox{where}\; q(L)=\int_{0}^{L}\sqrt{2m(x)}dx,\; \; n=0,\pm 1,\pm 2,....
\end{equation}
For a periodic BC we can take $\chi=0$, whereas for an anti-periodic BC we can consider $\chi=\pi$, which provides the allowed values of $\nu_n$ as follows.
\begin{equation}\label{nunBC}
\text{Periodic BC}:	\nu_n =i\pi n/q(L);\;\; \text{Anti-periodic BC}:	\nu_n =i\pi(n+1/2)/q(L).
\end{equation}
Note that, since the eigenvalues are purely imaginary, and discrete, the GIS constructed here are not a Glauber state in ordinary sense.
\subsection{Uncertainty measure}
 Let us revisit the covariance for our $\mathcal{P}\mathcal{T}$ symmetric scenario. 
 In the present situation, we can not rule out the possibility of correlation between $\hat{X}$ and $\hat{\Pi}$. In order to obtain the correct uncertainty relation, we need to invoke Heisenberg-Robertson or Schr\"{o}dinger uncertainty relation, which reads
 \begin{equation}\label{RSUP}
 	\Sigma+\frac{i}{2} \Omega\ge 0,
 \end{equation}
 where $\Sigma$ is the covariance matrix defined by $\Sigma_{ab}=\frac{1}{2}\langle \{X_a, X_b\}\rangle-\langle X_a\rangle \langle X_b\rangle$, and $\Omega$ is the symplectic matrix obtained from the commutation relations $\Omega_{ab}=-i\langle[X_a,X_b]\rangle_\eta$. In our case $a,b\in\{1,2\}$, with $X_1=\hat{X}$ and $X_2=\hat{\Pi}$. Therefore, our generalized deformed symplectic matrix is $\Omega= \left(\begin{array}{cc}
 	0 & \langle \theta\rangle \\
 	-\langle \theta\rangle & 0
 \end{array}\right)$. For the eigenstates of $\hat{A}_{-}$, we have
\begin{equation}
	\Sigma=\frac{1}{2}\langle \theta\rangle \mathbb{I}_2.
\end{equation}
The eigenvalues of $\Sigma+\frac{i}{2}\Omega$ are $\{0,\langle\theta\rangle\}$. 
The expecation value on the state $\vert\nu\rangle=\vert \psi\rangle_R$ of the operator $\mathcal{O}$ is defined by
\begin{equation}
	\langle \mathcal{O}\rangle_{\vert \nu\rangle}:= \frac{\langle \psi_L\vert \hat{\mathcal{O}}\vert \psi_R\rangle}{\langle \psi_L\vert  \psi_R\rangle} = \frac{\langle \psi_R\vert \hat{\eta} \hat{\mathcal{O}}\vert \psi_R\rangle}{\langle \psi_L\vert  \psi_R\rangle}.
	\end{equation}
One can always choose the normalization factor such that $\langle \psi_L\vert  \psi_R\rangle=1$.
 In the following, all expectation values are computed on the IS $\vert \nu\rangle$, unless otherwise specified. We consider here the normalized IS, i.e., $\langle \nu\vert \eta\vert \nu\rangle=1$. Then, we have
\begin{eqnarray}
\langle \hat{A}_{-}\rangle=\nu,\; \langle \hat{A}_{+}\rangle=\nu^*,\; \langle \hat{A}_{-}^2\rangle=\nu^2,\; \langle \hat{A}_{+}^2\rangle=(\nu^*)^2,\nonumber \\
\langle \hat{A}_{+}\hat{A}_{-}\rangle= \vert \nu\vert^2,\; \langle \hat{A}_{-}\hat{A}_{+}\rangle= \vert \nu\vert^2 + \langle \theta\rangle.
\end{eqnarray}
Consequently
\begin{eqnarray}
\langle \hat{X}\rangle &=&\sqrt{2}\nu_r,\; \langle \hat{\Pi}\rangle =\sqrt{2}\nu_c, \label{XPex}\\
\langle \hat{X}^2\rangle &=& 2\nu_r^2 +\frac{1}{2}\langle \theta\rangle,\;
\langle \hat{\Pi}^2\rangle =2\nu_c^2 +\frac{1}{2}\langle \theta\rangle.\label{XPsqex}
\end{eqnarray} 
Since we have $\{\hat{X},\hat{\Pi}\}=i(\hat{A}_{+}^2-\hat{A}_{-}^2)$, we also get
\begin{equation}
\frac{1}{2}\{\hat{X},\hat{\Pi}\}=2\nu_r\nu_c
\end{equation}
Thus, we have
\begin{eqnarray}
\Sigma_{11}=	(\Delta \hat{X})^2=\frac{1}{2}\langle \theta\rangle,\; \Sigma_{22}= (\Delta \hat{\Pi})^2=\frac{1}{2}\langle \theta\rangle,\; \Sigma_{12}=\Sigma_{21}=0.
\end{eqnarray}
Now $\Sigma +i\Omega$ has eigenvalues $\{0,\langle \theta\rangle\}$. In other words, $\Sigma +i\Omega$ is positive semidefinite for $\langle \theta\rangle>0$, which is a sufficient condition in order to $\Sigma$  be a bonafide covariance matrix. We shall show this for our toy model in subsequent section.\\
However, at present, we can conclude that the GIS $\vert \nu\rangle$ saturates the Schr\"{o}dinger-Robertson bound
\begin{equation}\label{vardef}
	\Delta \hat{X}\Delta \hat{\Pi} =\frac{1}{2}\langle\theta\rangle,\; \mbox{if}\; \langle \theta\rangle>0.
\end{equation}
From the uncertainty relation through the  $\theta(x)$, one can interpret $\theta$ as the position-dependent deformation function of the quadrature algebra. In other words, $\theta$ may be regarded as an effective uncertainty scale.
\section{Toy model: Effective mass $m(x)=m_0(1+\epsilon x^2)$}
In the toy model, we confine ourselves for $\beta_1=\beta_2=\beta$.
If we consider that the amplification parameter $\beta$ is small enough such that all the higher order terms of $\epsilon=-4m_0\beta$ can be regarded as negligible. We consider the mass profile 
\begin{equation}\label{massquad}
	m(x) = m_0(1+\epsilon x^2), \; \mbox{with}\; \epsilon\ge 0, \mbox{and}\;\mathcal{O}(\epsilon^2)\to 0.
\end{equation}
Note that since $\epsilon\ge 0$, the $\mathcal{P}\mathcal{T}$-symmetric parameter $\beta$ is negative.
If we use the PDEM ~\eqref{massquad} in ~\eqref{ricatti}, and put the ansatz
\begin{equation}
	\phi_{-}(x)=\phi_{0}(x)+\epsilon g(x),\; \mathcal{O}(\epsilon^2)\to 0,
\end{equation} 
the Ricatti equation~\eqref{ricatti} for the superpotential $\phi_{-}(x)$ provides  consistency conditions 
\begin{eqnarray}
	&& \phi_0'- \sqrt{2m_0} \phi_0^2 -4m_0\beta x \phi_0 = \sqrt{2m_0} (\lambda-\beta - V(x) ), \label{ricattiphi0} \\
&&	g' -\sqrt{2m_0}(2\phi_0 +2\sqrt{2m_0}\beta x)g = \frac{5}{4m_0}\sqrt{2m_0}+ \frac{1}{2}x^2 (\phi_0' + 4m_0\beta x \phi_0 -2\sqrt{2}\beta). \label{gprimeinphizero}
\end{eqnarray}
Recall that for constant mass Hamiltonian system $\phi_{-}$ is simply proportional to $x$. Keeping this in mind, we shall opt for the condition $\phi_{0}(x=0)=0$, which enables us to write the  solution of ~\eqref{ricattiphi0} for the potential ~\eqref{Vx}, as follows.
\begin{equation}
	\phi_0(x)= \frac{1}{\sqrt{2}} x \left[ -2\beta\sqrt{m_0} + k_{12} - \frac{(k_{12}+ \lambda_{12}) _1F^1[(5 + \lambda_{12}/k_{12})/4, 3/2, \sqrt{m_0}k_{12} x^2] }{_1F^1[ (1 + \lambda_{12}/k_{12})/4, 1/2,  \sqrt{m_0}k_{12} x^2]}\right],
\end{equation}
where $_1F^1(a,b; x)$ is Kummer's function or the Confluent Hypergeometric function, and 
\begin{equation}
	k_{12}=\sqrt{2k+4m_0\beta^2},\; \lambda_{12}=2\sqrt{m_0}(V_0-\lambda).
\end{equation}
Since we can tune the parameters through experimental means, let us choose the following values
\begin{equation}
	V_0=0,\; \lambda^2=(k+2m_0\beta^2)/(2m_0),
\end{equation}
which reduces the $\phi_0(x)$ to the simple form
\begin{equation}\label{phizero}
\phi_0(x)= \beta_\lambda x, \; \mbox{with}\; \beta_{\lambda}= \sqrt{2m_0}(\lambda -\beta).
\end{equation}
Using ~\eqref{phizero} in~\eqref{gprimeinphizero}, we obtain the first order linear equation of $g(x)$ 
\begin{eqnarray}
&&	g'-4m_0\lambda x g =\nu_0+\nu_2 x^2 +\nu_4 x^4, \label{gde}\\
&&	\mbox{with}\; \nu_0=\frac{5}{4m_0}\sqrt{2m_0},\; \nu_2 = \frac{\beta_\lambda}{2}-\sqrt{2}\beta,\; \nu_4 =2 m_0\beta\beta_\lambda.
\end{eqnarray}
By invoking the condition $g(0)=0$, we write the solution of ~\eqref{gde} as
\begin{equation}
	g(x)= -\nu_1 x -\nu_3 x^3 + \frac{1}{4}\sqrt{\frac{2\pi}{m_0\lambda}}(\nu_0+ \nu_1)  \mbox{Erf}[x \sqrt{2m_0\lambda}] e^{2m_0\lambda x^2}.
\end{equation}
 $\mbox{Erf}[z]$ is the integral of the Gaussian distribution, given by $\mbox{Erf}[z]=\frac{2}{\sqrt{\pi}}\int_0^z e^{-t^2}dt$, and the parameters
\begin{equation}
	\nu_1=\frac{1}{4m_0\lambda}(\nu_2+\frac{3\nu_4}{4m_0\lambda}),\; 
	\nu_3=\frac{\nu_4}{4m_0\lambda}.
\end{equation}
Now we have a complete expression for the superpotential $\phi_{1}=\phi_{-}(x)$ as
\begin{equation}
	\phi_{-}(x)=\beta_\lambda x + \epsilon g(x),
\end{equation}
Accordingly, the expression for $\phi_{+}(x)$ reads
\begin{equation}
	\phi_{+}(x)= \phi_{-}(x)+ 2\sqrt{2m_0}\beta x(1+\epsilon x^2/2).
\end{equation}
\begin{figure}
	\includegraphics[width=8cm]{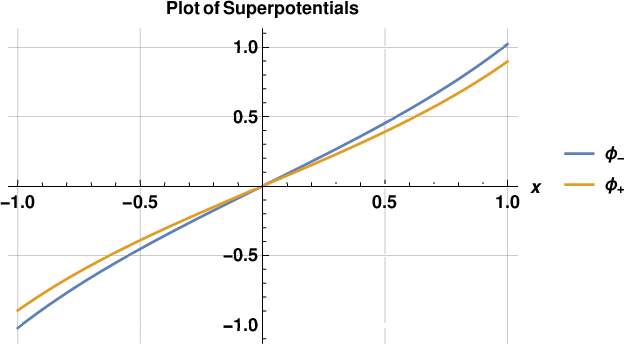}
	\caption{Superpotentials $\phi_{\pm}(x)$   in $x\in (-1,1)$ for the parameter values $m_0 = 1/2, \lambda= 1/2, \beta=- 1/16$, i.e., $\epsilon\to 1/8$.  }
	\label{phim}
\end{figure}
In figure FIG.~\ref{phim} we plot $\phi_{\pm}(x)$ for parameter values $m_0=1/2,  \beta=-1/16$ in the region $x\in(-1,1)$ (box length $=2$). Since around $x=0$, the $\mathcal{P}\mathcal{T}$-symmetric contribution vanishes, we see that $\phi_{\pm}$ coincides around the origin, as expected. In particular, the difefrence between two superpotentials $\phi_{\pm}$ is due to the $\mathcal{P}\mathcal{T}$-symmetric contribution. Interestingly, the nature of plot does not depend on the sign of $\beta$ (or equivalently the perturbation parameter $\epsilon$). 
Now we can write the noncommutative deformed parameter $\theta(x)$ as
\begin{equation}
	\theta(x)=2\lambda +\epsilon [ (3\beta-\lambda)x^2 +\frac{2}{\sqrt{2m_0}}g'].
\end{equation}
\begin{figure}
	\includegraphics[width=8cm]{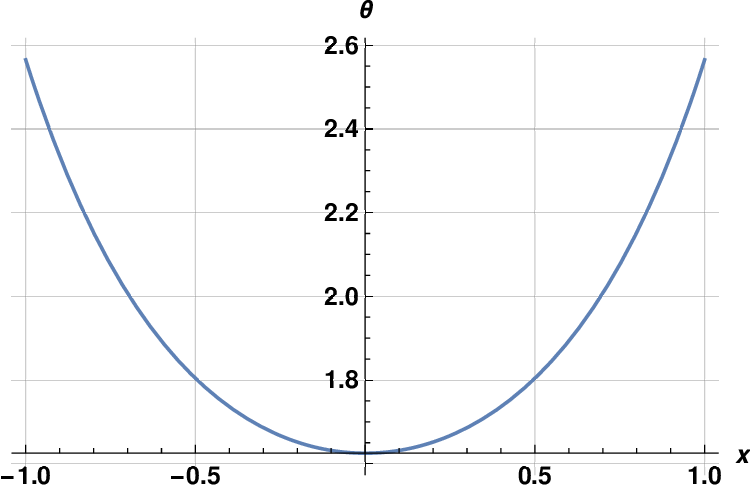}
	\caption{Deformation function ($\theta(x)$) of quadrature algebra,  in $x\in (-1,1)$ for the parameter values $m_0 = 1/2, \lambda= 1/2, \beta=- 1/16$, i.e., $\epsilon\to 1/8$.  }
	\label{thetax}
\end{figure}
In figure FIG.~\ref{thetax}, we plot $\theta(x)$ for parameter values $m_0=1/2,  \beta=-1/16$ in the region $x\in(-1,1)$ (box length $=2$).
Note that, for $\beta=0$ and $\epsilon=0$, we have $\theta=2\lambda$, which may be identified with the length scale  (similar to the Planck's constant $\hbar$). Thus $\lambda$ may be considered to $\lambda=1/2$. In particular, we can identify $\theta(x)$ as the position dependent effective Planck's constant, which reduces to the original Planck's constant for constant mass and Hermitian limit. From now on we shall consider $\lambda=1/2$.
To obtain the coherent state structure for our system, one can solve the right-eigenvalue equation of $\hat{A}_{-}$, which reads in position representation as follows
\begin{equation}\label{evaluexeqnr}
	\alpha(x)\frac{d}{dx}(\alpha(x)\psi_{r\xi}(x))+\phi_{-}(x)\psi_{r\xi}(x)=\xi \psi_{r\xi}(x),\;\mbox{with}\; \xi\in \mathbb{C},
\end{equation}
where $\psi_{r\xi}(x)=\langle x\vert \xi\rangle_r$. Without loss of generality, we shall consider $m_0=1/2$. From ~\eqref{psirnupsir0relation}, we see that the generalized coherent states $\psi_{r\xi}$ can be easily constructed from the vacuum $\psi_{r0}$.  The solution of ~\eqref{evaluexeqnr} for $\xi=0$ reads
\begin{eqnarray}\label{psizeror}
	 \psi_{r0}(x) =l_0 e^{-\frac{1}{4}x^2}[ 1+\frac{1}{4}\beta x^2(-1 + \frac{1}{2}x^2
	  + 11 _2F^2[\{1,1\},\{3/2,2\};x^2/2] )].
\end{eqnarray}
Here, $_2F^2[a,b;z]$ is the generalized hypergeometric function.
The normalization constant $l_0 \approx 1/\sqrt{1.56804}$. What follows is that the generalized coherent states are given by
\begin{equation}
	\psi_{r\xi}(x)=l_0 e^{\xi x-\frac{1}{4}x^2} [1+\frac{1}{4}\beta x^2 (-1-\frac{4}{3}\xi x + \frac{1}{2}x^2+ 11 _2F^2[\{1,1\},\{3/2,2\};x^2/2])] +\mathcal{O}(\beta^2).
\end{equation}
Using ~\eqref{nunBC} and ~\eqref{nunallowed}, we see that for the mass profile ~\eqref{massquad}, under the periodic BC ($\chi=0$), the allowed values of $\xi$ are
\begin{equation}\label{xin}
\xi_n= in\pi(1+\beta L^{2}/3)/L,\; n=0,\pm 1,\pm 2,.......,
\end{equation}
which enables us to index the GIS with integer as $\psi_{rn}$.
\begin{figure}
	\includegraphics[width=8cm]{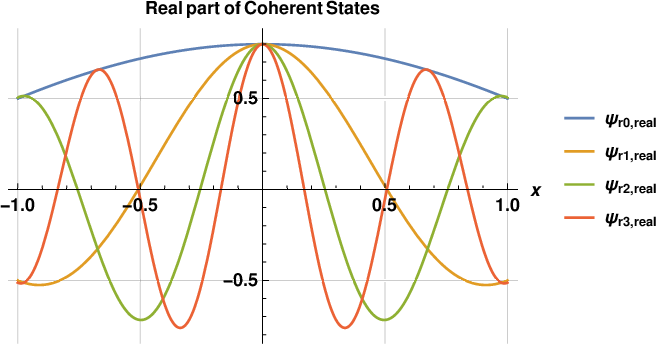}
	\caption{Real part $\Re(\psi_{rn})$ of the intelligent states $\psi_{rn}$. }
	\label{psireal}
\end{figure}
\begin{figure}
		\includegraphics[width=8cm]{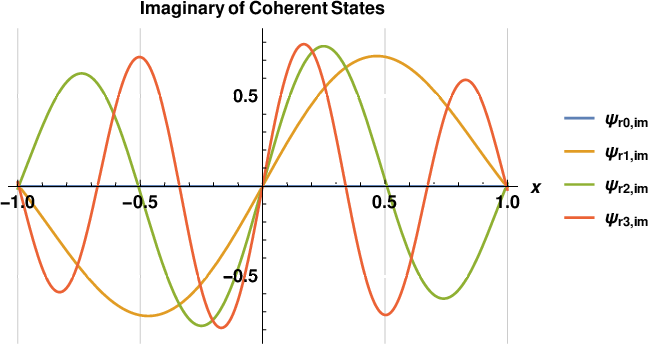}
	\caption{Imaginary part $\Im(\psi_{rn})$ of the intelligent states $\psi_{rn}$.}
	\label{psiim}
\end{figure}
In FIG.\ref{psireal}, we plot the real part of the GIS $\psi_{rn}$ for $n=0,1,2,3$, whereas the  FIG.\ref{psiim}, indicates the imaginary part of the GIS $\psi_{rn}$ for $n=0,1,2,3$.  Since $\xi_n$ are purely imaginary, we see that the probability density is independent of $n$. In particular,
\begin{equation}
	\rho(x)=\eta\vert\psi_{rn}\vert^2=\vert l_0\vert^2 \eta e^{-\frac{x^2}{2}}[1+\frac{1}{2}\beta x^2(-1+\frac{1}{2}x^2 + 11 _2F^2[\{1,1\},\{3/2,2\};x^2/2])].
\end{equation} 
\begin{figure}
	\includegraphics[width=8cm]{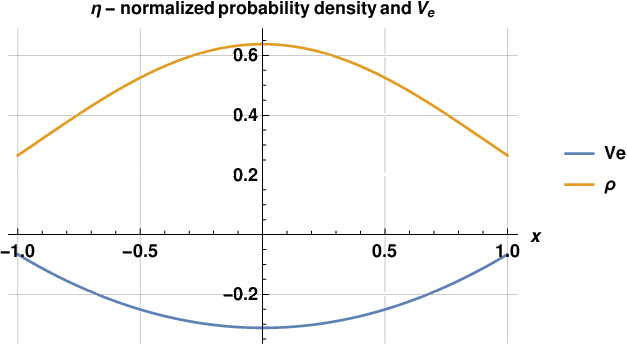}
	\caption{The density function $\rho(x)=\psi_{rn}^*\eta\psi_{rn}$, and the effective potential $V_{e}(x)$.}
	\label{rho}
\end{figure}
FIG.\ref{rho} indicates the $\eta$-noramalized density function $\rho(x)$ and the effective potential $V_e$. 
Using ~\eqref{XPex},~\eqref{XPsqex} and ~\eqref{xin}, we can write
\begin{eqnarray}
\langle \hat{X}\rangle_{\psi_{rn}} &=& 0,\; \langle \hat{\Pi}\rangle_{\psi_{rn}}=\frac{\sqrt{2}\pi n}{L}(1+\beta L^2/3),\\
\langle \hat{X}^2\rangle_{\psi_{rn}} &=& \frac{1}{2}\langle \theta\rangle_{\psi_{rn}},\; \langle \hat{\Pi}^2\rangle_{\psi_{rn}}=\frac{2\pi^2 n^2}{L^2}(1+\beta L^2/3)^2 +\frac{1}{2}\langle \theta\rangle_{\psi_{rn}}.
\end{eqnarray}
 Hence, from ~\eqref{vardef}, we conclude that, upto 1st order of $\beta$   
\begin{equation}
	\Delta \hat{X}\Delta\hat{\Pi}\vert_{\psi_{rn}}\ge \frac{1}{2}\langle \theta(x)\rangle_{\psi_{rn}}.
\end{equation}
For the chosen value of $\beta=-1/16$, the expectation value of $\theta(x)$ on any of the GIS $\psi_{rn}$ is given by
\begin{equation}
\langle \theta(x)\rangle = \int_{-1}^{1}\theta(x)\rho(x)dx\approx 1.93589 >0,
\end{equation}
which satisfy the requirement of the validity of~\eqref{vardef}.
\section{Conclusions}
We have explicitly demonstarted the deformed space structure induced by the  $\mathcal{P}\mathcal{T}$-symmetric position dependent effective mass (PDEM) system.  The generalized intelligent state structure and uncertainty principle is discussed explicitly with numerical computations. \\
The crucial observation from the present piece of article is that the uncertainty measure depends on the parameter $\beta$, which is responsible for $\mathcal{P}\mathcal{T}$-symmetric contribution. 
The limitation of the present work is that we have computed upto first order in $\beta$. Therefore, our estimation fails for strong $\mathcal{P}\mathcal{T}$-symmetric situation. For instance, for open quantum system, the interaction with the environment has to be sufficiently weak to match our theoretical results with experiments.  Moreover, because of the appearance of the terms containing $L$, at the denominator of the allowed values of the spectrum of $\hat{A}_{-}$ we should not take $L\to \infty$ limit in a straightforward manner to interpret the results for whole $\mathbb{R}$. In particular, the result stated here are valid only in a finite line segment $(-L,L)$.
\section{Acknowledgement}
Pinaki Patra is grateful to ANRF (erstwhile SERB) for financial support through project grant (File No: EEQ/2023/000784). 
\section{Author Declarations}
The authors have no conflicts to disclose.
\section{Availability of data}
Data sharing is not applicable to this article as no new data were created or analyzed in this study.

\end{document}